%% file: main.tex
\documentclass[pra,twocolumn,amsmath,amssymb,aps]{revtex4-2}
\makeatletter
\newcommand*{\rom}[1]{\expandafter\@slowromancap\romannumeral #1@}
\makeatother
\usepackage{graphicx}
\usepackage{dcolumn}
\usepackage{bm}
\usepackage[noend]{algpseudocode}
\usepackage[ruled,vlined]{algorithm2e}
\usepackage{amsmath,xparse,mleftright}
\usepackage{subfigure}
\usepackage{braket,color}
\usepackage{dsfont}
\usepackage[table]{xcolor}
\begin{document}

\title{Prime number factorization using a spinor Bose--Einstein condensate inspired topological quantum computer}%

\author{Emil Génetay Johansen}
\author{Tapio Simula}
\affiliation{Optical Sciences Centre, Swinburne University of Technology, Melbourne 3122, Australia\\
}

\begin{abstract}
Inspired by non-abelian vortex anyons in spinor Bose--Einstein condensates, we consider the quantum double $\mathcal{D}(\mathds{Q}_8)$ anyon model as a platform to carry out a particular instance of Shor's factorization algorithm. We provide the excitation spectrum, the fusion rules, and the braid group representation for this model, and design a circuit architecture that facilitates the computation. All necessary quantum gates, less one, can be compiled exactly for this hybrid topological quantum computer, and to achieve universality the last operation can be implemented in a non-topological fashion. To analyse the effect of decoherence on the computation, a noise model based on stochastic unitary rotations is considered. The computational potential of this quantum double anyon model is similar to that of the Majorana fermion based Ising anyon model, offering a complementary future platform for topological quantum computation.
\end{abstract}

\maketitle

\emph{Introduction:---}
The quest for fault-tolerant quantum computation is a substantial contemporary pursuit in science and technology \cite{2010qcqi.book.....N,1982IJTP...21..467F}. It is the intrinsic parallelism exhibited by quantum systems that is responsible for their extraordinary computational potential, and consequently, quantum systems could serve as an arena for simulating algorithms of exponential complexity. From an engineering point of view, the main hurdle in the construction of quantum devices is the mitigation of environmental noise that causes decoherence. The encoded  quantum information becomes distorted due to decoherence, which is why error correcting protocols \cite{2013qec..book.....L,2016Natur.540...44G,2009arXiv0904.2557G} are imperative for successful rectification of such distortions. However, error correcting schemes are generally very expensive to carry out. The idea of employing two dimensional systems that are intrinsically robust, such as systems exhibiting topological phases of matter \cite{1973JPhC....6.1181K,2009arXiv0904.2771K,2017RvMP...89d1004W}, has led to a new paradigm in quantum computing known as \emph{topological quantum computation} (TQC) \cite{2012itqc.book.....P,2018QS&T....3d5004F,2017ScPP....3...21L,2017arXiv170506206R,2009arXiv0904.2771K}. 

Such topologically ordered systems may be inhabited by special kinds of quasiparticles called \emph{non-abelian anyons} \cite{1977NCimB..37....1L,2008RvMP...80.1083N,1990SDCMP..10.....W}, which may pave the way towards the realization of TQC. When non-abelian anyons are exchanged, their wave-function transforms according to representations of the \emph{braid group}. This is in contrast to bosonic and fermionic wavefunctions, which transform rather trivially under the action of the permutation group. The anyonic quantum states are subject to topologically protected unitary transformations when braiding of their worldlines is performed. Such braiding of anyons serves as a possible way to implement topologically protected quantum circuits. 

Notable anyon models include the Fibonacci and Ising anyons, which both belong to the family of $\rm{ SU}(2)_k$ models that are based on non-abelian Chern--Simons theory \cite{2018QS&T....3d5004F,1989CMaPh.121..351W,PRXQuantum.2.010334,1998NuPhB.516..704F,1999tald.conf..177D}. All ${\rm{SU}(2)}_k$ models are universal except for the cases $k=2$ and $k=4$, the former of which could potentially be realised by Majorana fermion zero mode quasiparticles \cite{sarma2015majorana}. 

Here we consider another class of anyon models known as \emph{quantum doubles} \cite{1995hep.th...11201D,2009arXiv0904.2771K,Gould1993QuantumDF,2009NJPh...11e3009B,2006JMP....47j3511D}. Specifically, we are focusing on the quantum double of the quaternion group $\mathcal{D}(\mathds{Q}_8)$, inspired by its connection to the non-abelian vortex anyons in spinor Bose--Einstein condensates \cite{2019PhRvL.123n0404M,2010arXiv1001.2072K,2008PhRvL.100r0403K,2010PThPS.186..455K,2019qvht.book.....S}. In particular, the unbroken high-temperature phase of an $F=2$ spinor BEC may, through spontaneous symmetry breaking, collapse to the biaxial nematic phase \cite{2010arXiv1001.2072K,RevModPhys.85.1191} characterised by the binary dihedral-4 group ${D}^*_4$. It is conceivable that this may further be broken down to its ${D}^*_2$ subgroup, which is isomorphic to the quaternion group $\mathds{Q}_8$, considered here. The $\mathds{Q}_8$ is a particularly small subgroup representing little residual symmetry, which should be beneficial for the prospect of its experimental realizability. The $\mathds{Q}_8$ based quantum double model has previously been considered in \cite{2012NJPh...14c5024B} albeit in a different context. 

In the present work, we are turning our focus to applications within quantum double TQC. We derive explicitly all pertinent information, such as particle content, fusion rules and braiding rules, for the non-abelian quantum double model $\mathcal{D}(\mathds{Q}_8)$. We also design qubit structures (fusion trees), which are exploiting the full computational power of the model to optimize its utility. As a proof of principle demonstration, we then design and compile a quantum circuit architecture which allows us to factorize the number 15 into its prime number constituents using Shor's algorithm \cite{1995quant.ph..8027S}. Experimental realizations of Shor's algorithm using other platforms have been studied in \cite{2001Natur.414..883V,2007PhRvL..99y0504L,2007PhRvL..99y0505L,2009Sci...325.1221P,2012NaPho...6..773M,2012NatPh...8..719L,2016Sci...351.1068M,2019PhRvA.100a2305A,2020OExpr..2818917D}.  

Quantum doubles based on discrete gauge groups emerge through the Higg's mechanism by breaking particular symmetries of the initial Yang--Mills--Higg's Lagrangian \cite{1995hep.th...11201D}. Due to the resulting discrete structure, braiding the anyons of the theory can only implement a finite set of unitary transformations, implying that the corresponding braid group is non-universal. 
To remedy this, measurement based fusion protocols \cite{2015PhRvA..92a2301L,2003PhRvA..67b2315M} could be implemented, allowing one to carry out phase gate rotations of arbitrary angle. However, universality can only be achieved this way at the expense of sacrificing some fault-tolerance. 

In the specific quantum circuit considered here, only one additional noisy gate is required, as all of the other gates can be implemented exactly within the anyon model by braiding alone. We employ a noise model based on stochastic unitary rotations to study its effect on the computational process. We also account for the error accumulation in the form of leakage from the computational subspace to its complement \cite{2003PhRvA..67b2315M}. 

\emph{Quantum double of the quaternions:---}
Consider a condensate with spinor degrees of freedom governed by the (2+1)-dimensional non-abelian Yang--Mills--Higgs action 
\[S= \int_{\mathds{R}^{2+1}} d^3 x (F_{\mu \nu}^a F^{\mu \nu}_a + (\mathcal{D}^{\mu} \phi)^{\dagger} \mathcal{D}^{\mu} \phi - V(\phi))\]
with $\rm{SU}(2)$ symmetry, where $F_{\mu \nu}^a$ is the gauge curvature tensor, $\phi$ is the Higgs field, $\mathcal{D}$ is the covariant derivative and $V$ is the potential. We envisage the system cooling down and undergoing a symmetry breaking process to a subgroup $H \subset {\rm SU}(2)$. Here we consider an ordered phase corresponding to $H = \mathds{Q}_8$, which means that the pertinent excitations are determined by the homotopy theory of $\mathcal{D}(\mathds{Q}_8)$. The group $\mathds{Q}_8$ has eight elements and five conjugacy classes according to the partitioning $\mathds{Q}_8 = \{ \langle e \rangle,\langle \bar{e} \rangle,\langle i,\bar{i} \rangle,\langle j,\bar{j} \rangle,\langle k,\bar{k} \rangle\}$, where $e$ is the identity and the bar denotes conjugation. The quantum double of a finite group is an algebraic construction that simultaneously involves the group and its Fourier dual \cite{1999JPhA...32.8539K}. The particle content of the quantum double is consequently defined by the irreducible representations of this algebra. In particular, the possible species of one type of particle, referred to as \emph{fluxons}, are categorised according to the conjugacy classes of $\mathds{Q}_8$. Moreover, a second particle type, known as $chargeons$, also exist in the excitation spectrum, which inhabit the reciprocal space of $\mathds{R}^{2+1}$ and are defined by the irreducible representations of $\mathds{Q}_8$. Hence, the chargeons and fluxons are related by a generalised Fourier transform, which establishes a particle-vortex duality in the model. These two particle types can also coexist, thus forming composite objects known as $dyons$, under the condition that the chargeon group element commutes with the fluxon one, meaning that a $dyon$, denoted by $(C,\Gamma (Z_C))$, is specifically defined by a conjugacy class $C$ and an irreducible representation of its centralizer $\Gamma (Z_C)$. The centralizers of each conjugacy class of $\mathds{Q}_8$ are listed in Table \ref{centralizer}.

\begin{table}[!b]
\caption{Conjugacy classes $C$ of $\mathds{Q}_8$ and their corresponding centralizers $Z(C)$.}
\begin{ruledtabular}
\begin{tabular}{ll}
\textit{Conjugacy classes}& \textit{Centralizers}\\
$C_e = \langle e \rangle$ & $Z(C_e) = \mathds{Q}_8$\\
$C_{\bar{e}} = \langle \bar{e} \rangle$ & $Z(C_{\bar{e}}) =\mathds{Q}_8$\\
$C_i = \langle i, \bar{i} \rangle$ & $Z(C_i) = \mathds{Z}_4 = \{e,\bar{e}, i, \bar{i}\}$\\
$C_j = \langle j, \bar{j} \rangle$ & $Z(C_j) = \mathds{Z}_4 = \{e,\bar{e}, j, \bar{j}\}$\\
$C_k = \langle k, \bar{k} \rangle$ & $Z(C_k) = \mathds{Z}_4 = \{e,\bar{e}, k, \bar{k}\}$
\end{tabular}
\label{centralizer}
\end{ruledtabular}
\end{table}

The group $\mathds{Q}_8$ has four one-dimensional irreducible representations comprising one trivial $\Lambda_0$ and three non-trivial ones $\Lambda_a$ ($a=1,2,3$), in addition to one two-dimensional representation $\Lambda_4$. The remaining centralizers have four one-dimensional irreducible representations given by one trivial, $\Pi_0$, and three non-trivial ones $\Pi_a$ ($a=1,2,3$), which are simply permutations of one another. In total, $\mathcal{D}(\mathds{Q}_8)$ has 22 particle species, comprising four pure fluxons
\begin{equation}
    \mathds{\bar{1}} = (\bar{e},\Lambda_0) \hspace{5mm}{\rm and}\hspace{5mm}  \Phi_x = (C_x,\Pi_0),
\end{equation}
where $x=i,j,k$, four pure chargeons
\begin{equation}
    \rho_y = (e,\Lambda_y) \hspace{5mm}{\rm and}\hspace{5mm}  \Delta = (e,\Lambda_4),
\end{equation}
where $y=1,2,3$, and 14 composite dyons
\begin{align}
    \tilde{\Phi}_x &= (C_x,\Pi_2),\;\;\;  \bar{\rho}_y = (\bar{e},\Lambda_y),\;\;\; \bar{\Delta}_4 = (\bar{e},\Lambda_4), \notag \\ \Sigma_x &= (C_x,\Pi_1) \hspace{5mm}{\rm and}\hspace{5mm} \tilde{\Sigma}_x = (C_x,\Pi_3).
\end{align}
In addition to these particles, the pure vacuum sector is denoted by $\mathds{1} = (e,\Lambda_0)$.

\emph{Fusion and braiding:---}
When two non-abelian anyons are fused their joint tensor representation branches into its irreducible orthogonal blocks, which correspond to the possible particle outcomes of the fusion. The particle types that emerge from the decomposition can be conveniently obtained using the so called Verlinde's formula \cite{1988NuPhB.300..360V}
\begin{equation}
    N_{AB\gamma}^{C\alpha \beta} = \sum_{D,\delta} \frac{S_{AD}^{\alpha \delta} S_{BD}^{\beta \delta} S_{CD}^{\gamma \delta}}{S_{eD}^{0 \delta}},
\end{equation}
where $A,B,C,$ and $D$ denote conjugacy classes and $\alpha,\beta,\gamma,\delta$ label the centralizer irreducible representations. The explicit form of the modular $S$-matrix is provided in Supplemental Material \cite{supplement}. The complete set of fusion rules are also listed in \cite{supplement}, and a subset of these
\begin{align}
\Phi_x \otimes \Sigma_x &= \Delta \oplus \bar{\Delta}, \; \Phi_x \otimes \Sigma_y = \Phi_z \oplus \tilde{\Phi}_z,\\
    \Phi_x \otimes \Phi_x &= \mathds{1} \oplus \mathds{\bar{1}} \oplus \rho_x \oplus \bar{\rho}_x, \; \Phi_x \otimes \Phi_y = \Phi_z \oplus \tilde{\Phi}_z,\notag\\
    \Sigma_x \otimes \Sigma_x &= \mathds{1}  \oplus \rho_x \oplus \bar{\rho}_y \oplus \bar{\rho}_z, \; \Sigma_x \otimes \Sigma_y = \Sigma_z \oplus \tilde{\Sigma}_z,\notag
\end{align}
  are required for defining our qubit Hilbert spaces.
 
\emph{Computational universality:---}
Several different anyon systems would qualify as a qubit architecture. For the purpose of demonstrating Shor's algorithm it would make sense to design our TQC model such that its computational power is maximized. 
Since the specific proof of concept objective is to factor the number 15, it is tempting to base our Hilbert space on either $\Phi_x$ anyons or $\Sigma_x$ anyons as both of these have four fusion outcomes, which means that only two such qudits would be required to represent the numbers from 1 to 16. However, by analyzing the topology of the resulting Hilbert space we find that universality of the model will become a major consideration.

\emph{Hopf-fibrations:---}
A four-level system transforms under $\mathrm{SU(4)}$ and since this group is acting on a space with a total of $2 \cdot 4 = 8$ dimensions, the spherical surface that is being rotated is $8-1=7$ dimensional, that is, a 7-sphere $S^7$. Further, it follows from Adam's theorem \cite{Adams60onthe} that topological spheres of dimension $0,1,3,$ and $7$ have the local structure of a fiber bundle, thus allowing us to decompose the manifold into its base space and a fiber such that $f: S^d \longrightarrow S^n \times S^m$, where $d=n+m$. Such a map $f$ is known as a Hopf fibration \cite{Lyons2003AnEI,mosseri_ribeiro_2007} and when applied to the four-level system, it locally maps the manifold $f: S^7 \longrightarrow S^4 \times S^3$, where $S^4$ is the base space and $S^3$ is the fiber. We can apply this map iteratively, which allows us to further decompose $S^3$ according to $f: S^3 \longrightarrow S^2 \times S^1$, that is a regular 2-sphere and a circle, from which we can conclude that $S^7 \simeq S^4 \times S^2 \times S^1$, locally. 

Since we consider these maps in the context of a quantum mechanical system, the $S^1$ degree of freedom pertains to the ${\rm U}(1)$ gauge  freedom, which is an experimentally unmeasurable symmetry of the amplitude. We may thus consider the projective Hilbert space, meaning that the effective topology of the manifold is $S^4 \times S^2$. Since computational universality entails that we must be able to generate a topologically dense cover over the manifold, we conclude that achieving this is much harder for a 4-level qudit than a 2-level qubit. Specifically, since qubits transform according to ${\rm SU}(2)$, which rotates a $2\cdot 2 -1=3$ dimensional sphere $S^3$, the effective topology is $S^2$ (the Bloch sphere) due to its local gauge fiber structure. Moreover, if we define a stereographic projection of $S^2$ onto $\mathds{R}^2$ through the map $s: S^2 \longrightarrow \mathds{R}^2 \cup \infty$, we may conclude that in order to cover $S^2$ densely, we need to find a two-dimensional basis and make sure that we have elements of infinite order in the braid group. For a non-universal three-stranded braid group $\mathds{B}_3$, this can be achieved by supplementing the generator set with an irrational phase gate \cite{PRXQuantum.2.010334}. Note that for a two qubit system, which also has four levels, we have 6 anyons and thus a six-stranded braid group $\mathds{B}_6$, which has five generators, whereas a single 4-level qudit still only transforms under $\mathds{B}_3$ with two generators. Consequently, it is much harder to span the complicated 4-level sphere in the qudit case due to the fewer number of generators and as a result, a less powerful braid group.

\emph{Circuit architecture:---}
We have arrived at the conclusion that basing our quantum circuit on anyons of the same species probably would make it difficult to implement the logic gates required in Shor's algorithm, since such systems have four levels. Moreover, calculating the braid group generators shows that the resulting group is either trivial or close to trivial. This leads to a conjecture that diversifying the qubit architecture might be a good approach for maximizing the computational power of the anyon model. Specifically, basing the individual qubits on either $\Sigma_x$ or $\Phi_y$ type anyons (where $x,y = 1,2,3$), or a mixture of the two, yields a particularly strong model as the resulting braid group order is maximized and simultaneously the number of non-computational basis states will be minimized, reducing the expected leakage into these states.

\begin{figure}[!]
    \centering
    \includegraphics[width=1\columnwidth]{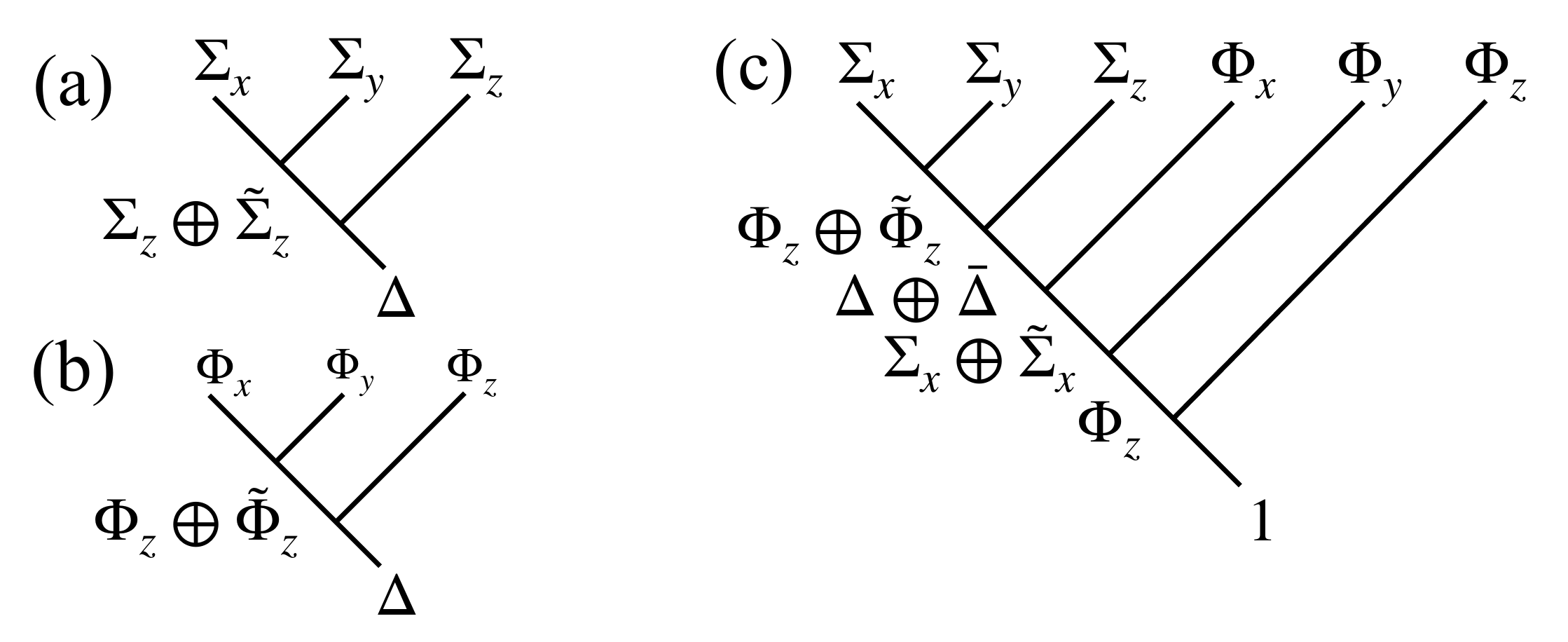}
    \caption{(a) Qubit based on $\Sigma$ anyons. (b) Qubit based on $\Phi$ anyons. (c) Two qubit anyon system based on $\Sigma$ and $\Phi$ anyons.
    }
    \label{qubit}
\end{figure}

\emph{$\Sigma \Phi$ anyon computer:---}
To implement the circuit illustrated in Fig.~S1 \cite{supplement}, four qubits are required. This can be achieved by defining three of the qubits as in Fig.~\ref{qubit} (a) and the last one as in Fig.~\ref{qubit} (b), where the controlled operations are implemented between qubits of the former kind with those of the latter.
Any two qubit interaction in the circuit will thus be of the form presented in Fig.~\ref{qubit} (c) where all anyons are distinguishable. Note that all vertices are independent meaning that the total two qubit Hilbert space is $2^3 \cdot 1 = 8$ dimensional, which implies that we have four non-computational states in addition to the four computational ones defined by $\mathcal{H}_{\rm comp} = {\rm span}\{ \ket{\Phi_z,\Delta,\Sigma_x},\ket{\tilde{\Phi}_z,\Delta,\Sigma_x},\ket{\Phi_z,\Delta,\tilde{\Sigma}_x }, \ket{\tilde{\Phi}_z,\Delta,\tilde{\Sigma}_x } \}$. The single qubit braid matrices $\sigma_1$ and $\sigma_2$ are 
\begin{align}
  &  \sigma_1^{ (\Phi_x \Phi_y)} = \begin{pmatrix}
    -1 & 0\\
    0 & e^{i\frac{\pi}{2}}
    \end{pmatrix},\  
    \sigma_2 ^{(\Phi_x \Phi_y)} = \frac{1}{\sqrt{2}}\begin{pmatrix}
    e^{i\frac{3 \pi}{4}} & e^{i\frac{5 \pi}{4}}\\
    e^{i\frac{5 \pi}{4}} & e^{i\frac{3 \pi}{4}}
\end{pmatrix}\notag\\
 &   \sigma_1 ^{(\Sigma_x \Sigma_y)} = \begin{pmatrix}
    e^{-i\frac{\pi}{4}} & 0\\
    0 & e^{-i\frac{3 \pi}{4}}
    \end{pmatrix},\  
    \sigma_2 ^{(\Sigma_x \Sigma_y)} = \frac{1}{\sqrt{2}} \begin{pmatrix}
    -1 & 1\\
    1 & -e^{i\frac{\pi}{2}}
\end{pmatrix}\notag\\
 &   \sigma_1^{ (\Sigma_x \Phi_y)} = \begin{pmatrix}
    -e^{-i\frac{\pi}{4}} & 0\\
    0 & e^{i\frac{\pi}{4}}
    \end{pmatrix},\  
    \sigma_2 ^{(\Sigma_x \Phi_y)} = \frac{1}{\sqrt{2}} \begin{pmatrix}
    1 & -1\\
    -1 & e^{i\frac{\pi}{2}}
\end{pmatrix},\notag
\end{align}
which can be derived with the aid of the Supplemental Material \cite{supplement}. The two qubit braids \cite{supplement} can be obtained by means of graphical calculus given the information contained in the single qubit ones \cite{2018QS&T....3d5004F}. We proceed by making a few pertinent remarks. For a given $i$ the braid matrices $\sigma_i$ are equivalent up to a global phase factor, which implies that they have the same effective projective action. Remarkably, they also map projectively onto the Ising anyon braid matrices given by $\rm{SU}(2)_2$ Chern--Simons theory. However, it is well known that the Ising anyon braids implement the Clifford group exactly, which is spanned by the Pauli matrices that form a representation of the quaternions. The $8\times 8$ two qubit braid matrices in the six anyon encoding scheme are provided explicitly in \cite{supplement}, and similarly one can prove that these map projectively onto the two qubit Ising anyon braid matrices, but in the eight anyon encoding scheme. Interestingly, many of the standard logic gates can be implemented exactly within this model, despite it being non-universal. For instance the Hadamard $\mathrm H$ and $\mathrm{CNOT}$ gates are
$   {\mathrm H} = \sigma_2 \sigma_1 \sigma_2 $, 
and
$    {\rm CNOT} 
    = \mathcal{P} \sigma_3^{-1} \sigma_4^{-1} \sigma_5^{-1} \mathcal{P} \sigma_3 \sigma_4 \mathcal{P} \sigma_3 \sigma_1$,
where $\mathcal{P}$ is a projection operator that can be regarded as a map $\mathcal{P}: \ \ \mathcal{H}_{\rm full} \longrightarrow \mathcal{H}_{\rm comp}$  projecting the full two qubit Hilbert space $\mathcal{H}_{\rm full}$ onto the computational subspace $\mathcal{H}_{\rm comp}$, thus containing all of the amplitude in $\mathcal{H}_{\rm comp}$. We provide the exact compiled forms of the $S$-gate, the Pauli-$X$, the Pauli-$Y$, the Pauli-$Z$ and the controlled-Pauli-$Z$ in Supplemental Material \cite{supplement}. Note that the CNOT is four dimensional while the two qubit braid matrices are eight dimensional. This means that amplitude will leak into the non-computational states when $\sigma_3$ is applied since this gate is the only one that couples the two subspaces. However, projection methods have been developed to manage the leakage, which, if successfully implemented, will have the effect of only braiding within the computational space. There also exist a subset of controlled two qubit braids known as weaves, which naturally cause very little leakage \cite{PhysRevB.75.165310}. However, this weaving method is only useful when one has a vacuum sector in the fusion product and when the model is universal. Here we instead suggest to perform a projective measurement $\mathcal{P}$, after each $\sigma_3$ braid. 

\begin{figure}[!]
    \centering
    \includegraphics[width=\columnwidth]{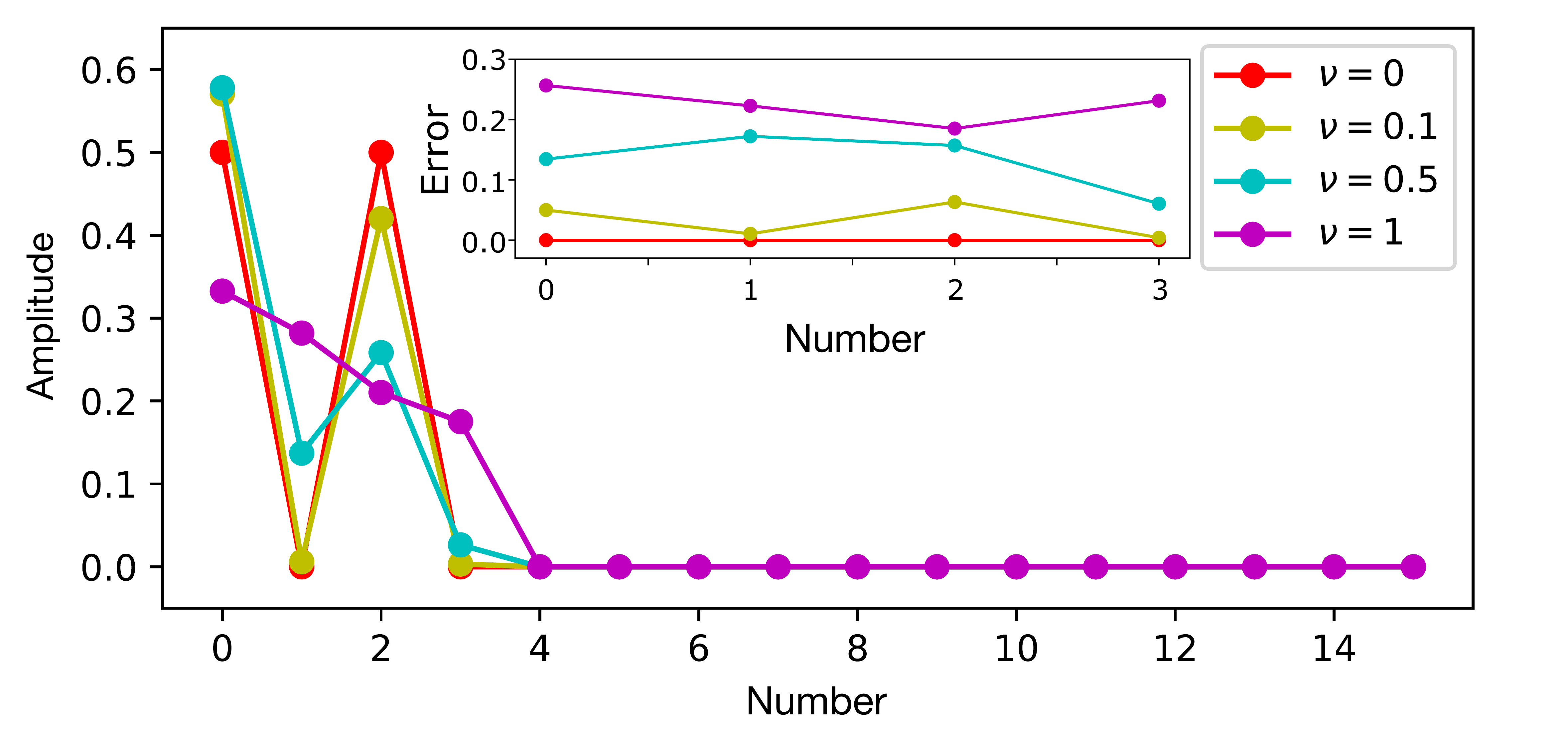}
    \caption{Prime number factorisation of 15 using Shor's algoritm. Amplitudes of the resulting superposition when four different noise levels are applied. The inset represents the statistical error corresponding to the first four data points, $0,1,2,3$, with non-zero amplitude. 
    }
    \label{results}
\end{figure}

As noted, the Hadamard and the CNOT can be implemented without any compilation error, given that the leakage error correction is carried out for the CNOT, so the only gate required for the purposes of our demonstration that cannot be implemented by means of braiding alone is the controlled-$\pi/2$. To implement this gate we suggest using similar scheme as developed in \cite{2015PhRvA..92a2301L}, where a reservoir of ancillary qubits are used to set up product states $\ket{\Psi} \ket{R_{\varphi/2}}$, where $\ket{R_{\varphi/2}}$ is phase rotated by an angle $\varphi/2$, from which the phase $R_{\varphi/2}$ can be extracted. However, such a measurement protocol is susceptible to noise and therefore in the results presented in Fig.~\ref{results} we have applied stochastic unitary rotations to simulate the effect of conventional noise on the computation. The rotational angles of arbitrary elements $U\in {\rm U}(4)$ are sampled from a normal distribution $N(0,\nu)$ with zero mean and variable standard deviation $\nu$, which can be interpreted as the noise strength \cite{PRXQuantum.2.010334}. Assuming that this can be successfully achieved with $\varphi = \pi$, all of the logical operations required for the implementation of the Shor's algorithm quantum circuit are available.

\emph{Factorisation of 15:---}
The result of the simulation of Shor's algorithm corresponding to the instance $N=15$ and $a=11$, using our $\mathcal{D}(\mathds{Q}_8)$ topological quantum computer simulator is shown in  Fig.~\ref{results}. Four different levels of noise corresponding to $\nu = 0,0.1,0.5,1$ are applied to the controlled-$\pi/2$ gate, which could not be realized by braiding alone. Figure~\ref{results} presents the probability distribution of the final state, showing two peaks with 50\% amplitude each, representing the numbers 0 and 2, when no noise is applied (red curve). The trivial number 0 is a false solution but measuring 2 solves the problem as the period can be computed as $r=\frac{2^2}{2}=2$, which yields the prime factors $gcd(a^{\frac{r}{2}}\pm 1, N)=gcd(11^{\frac{2}{2}}\pm 1, 15)=3,5$, where $a=11$ is chosen. Furthermore, the peaks become less distinct when the noise level is increased, eventually destroying the computation as the amplitude becomes too spread out. Each of the curves represent an average over 1000 realizations.

\emph{Conclusions:---}
 We have presented a model of a topological quantum computer based on the quantum double of the quaternions $\mathcal{D}(\mathds{Q}_8)$ inspired by its structural similarity to the superfluid phase that supports fractional vortices in spinor Bose--Einstein condensates as its fluxons \cite{2019PhRvL.123n0404M}. All pertinent information of the quantum double such as particle content, fusion rules and braiding rules were derived and a qubit architecture was designed to facilitate topological quantum computation. We performed a technology demonstration of this anyon model by carrying out prime number factorisation using Shor's algorithm. The recipe of the quantum double based TQC is generic and can be applied to any laboratory superfluid having a stable ground state symmetry characterised by a discrete non-abelian gauge group, whose topological excitations include non-abelian vortex anyons.

\begin{acknowledgements}
We are grateful to Joost Slingerland for generously sharing time to discuss algebraic aspects of the quantum double construction. This research was funded by the Australian Government through the Australian Research Council (ARC) Future Fellowship project FT180100020.
\end{acknowledgements}

\input{output.bbl}

\begin{appendix}
\section{Shor's algorithm}

Shor's algorithm consists of two main parts, a quantum step followed by a classical step. The algorithm is initiated by setting up disentangled product state of two registers $\ket{\psi} = \ket{0}^{\otimes n} \otimes \ket{0}^{\otimes n}$ of $n = 2 \cdot \lceil \log_2 (N)\rceil$ qubits, where the brackets denote the \emph{ceil} function that is rounding up the number to the closest integer and $N$ is the number being factorised. The factor two comes from the fact that two registers are required, one in which the integers $1,2,..,N$ are encoded and one which serves as a target when the controlled gates in the modular exponentiation function (MEF) is applied. The first register is then set up in an equal weight superposition by applying the Hadamard gate $H$ to all of the $n$ qubits in the register which results in a state 

\begin{equation}
 \ket{\Psi} = H^{\otimes n} \otimes I^{\otimes n} \ket{\psi}= \frac{1}{2^{n/2}} \left[\sum_{m=0}^{2^n-1} \ket{m}\right] \otimes \ket{0}^{\otimes n}.    
\end{equation}

\begin{figure}[!]
    \centering
    \includegraphics[width=\columnwidth]{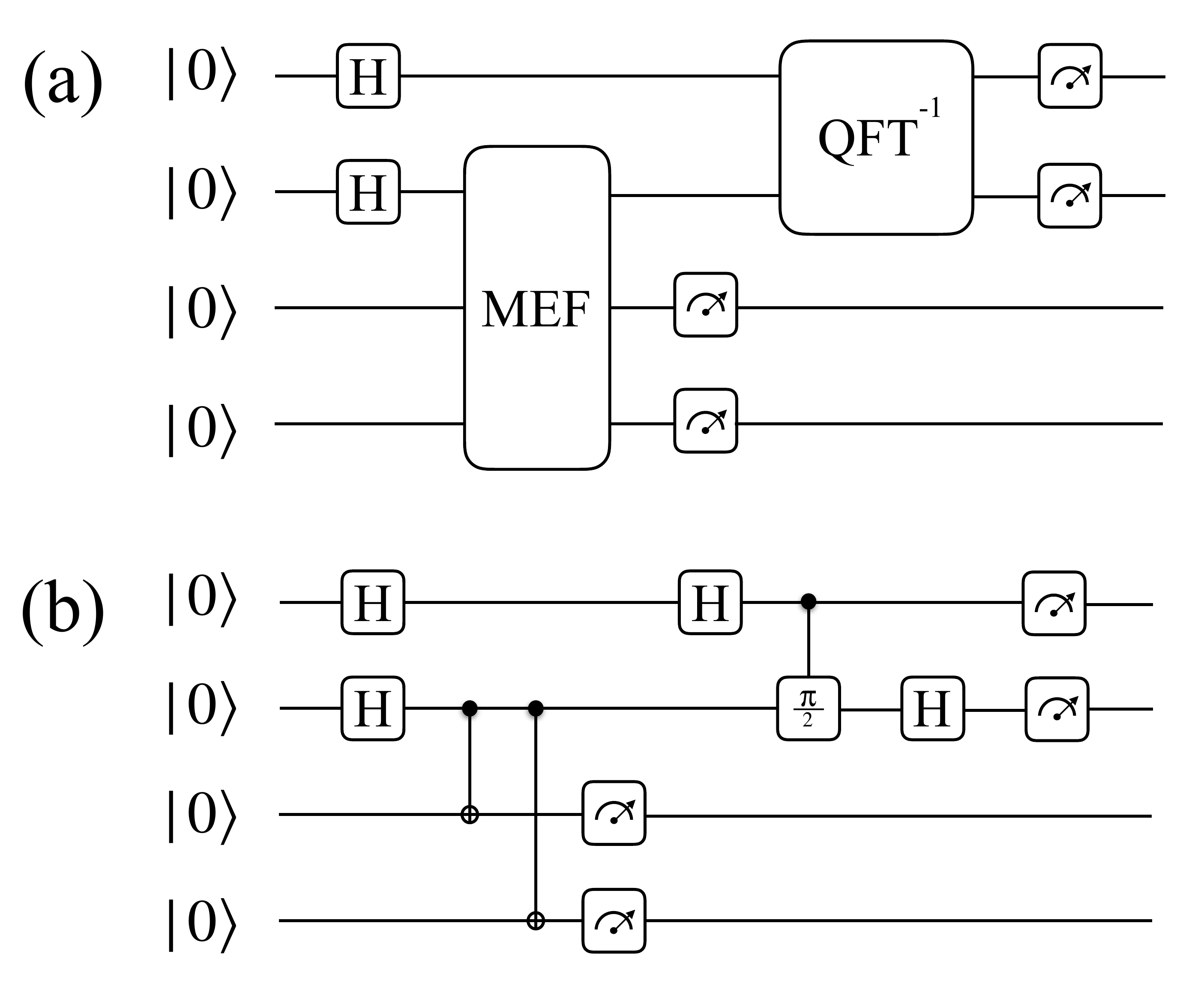}
    \caption{(a) Modular circuit for Shor's algorithm for $N=15$ and $a=11$. (b) Same circuit as in (a) but with the subroutines decomposed into elementary gate operations.
    }
    \label{circuit}
\end{figure}

Next, the quantum period finding subroutine is carried out on the full register which finds the period of the function $f(x) = a^x \ \ (mod \ \ N)$, where $a$ is an integer in the interval $1 < a < N$. This part truly is at the heart of Shor's algorithm as such a problem is inherently exponential in nature and cannot be solved efficiently by means of any classical analog. Quantum period finding can be further decomposed into two parts. First, the MEF is applied to the lower register resulting in 
\begin{equation}
{\rm MEF:} \ \ \ket{\Psi_{MEF}} = \frac{1}{2^{n/2}} \sum_{x=0}^{2^n-1}\ket{x} \otimes \ket{a^x \ \ (mod \ \ N)}, 
\end{equation}
where after the lower register is measured, thus projecting the full Hilbert space onto a subspace spanned by the states $\ket{x'}$ resulting in the same number $a^x \ \ (mod \ \ N)$. The last step before the final measurement is to apply the inverse quantum Fourier transform ${\rm QFT}^{\dagger}$ to the top register
\begin{equation}
{\rm QFT}^{\dagger}: \ \ \ket{\tilde{\Psi}_{{\rm MEF}}} = \frac{1}{2^{n/2}} \sum_{y=0}^{2^n-1} \sum_{x'} e^{-i 2\pi \frac{yx'}{2^n}}\ket{x'}, 
\end{equation}
which has the effect of destructively interfering the false solutions and constructively interfering the true solutions, resulting in sharp amplitude peaks pertaining to the states that solve the problem. One of these solution candidates is measured in the very last step. Suppose that a state $\ket{m}$ was measured. Then the rest of the algorithm can be completed classically as we only have to compute $gcd(a^{\frac{r}{2}} \pm 1, N)$, where the period $r$ can be obtained from $m = j \frac{2^n}{r}$, where $j$ is the smallest integer such that the equation is satisfied. However, in this work we are merely interested in a proof of concept demonstration of factorizing the number $15$ and if we pick $a=11$, the circuit can be reduced so that only two qubits are required in each register, instead of four. This is due to the fact that the MEF will always return only two states $\ket{1}$ and $\ket{11}$ for this particular instance of $a$. In Fig.~\ref{circuit} (a) the circuit is represented in its higher level modular form and in Fig.~\ref{circuit} (b) the different oracles are broken down into the elementary gates.

\section{Structure of the $\mathcal{D} (\mathds{Q}_8)$ anyon model}
\label{sec:level1}

Here we outline the structure of the $\mathcal{D} (\mathds{Q}_8)$ anyon model which is based on the quaternion group $\mathds{Q}_8$.

\subsection{Cayley table of the quaternion group $\mathds{Q}_8$}
Table \ref{table} shows the Cayley table for the quaternion group $\mathds{Q}_8$. The colors correspond to the five conjugacy classes of this group with eight group elements.

\begin{table}[t]
\caption{Cayley table of the quaternion group $\mathds{Q}_8$.}
\setlength\tabcolsep{0pt}
\centering
\begin{tabular}{  c | c | c | c | c | c | c | c | c | }
  \multicolumn{1}{c}{$\times$} & \multicolumn{1}{c}{$e$} & \multicolumn{1}{c}{$\bar{e}$} & \multicolumn{1}{c}{$i$} & \multicolumn{1}{c}{$\bar{i}$} & \multicolumn{1}{c}{$j$} & \multicolumn{1}{c}{$\bar{j}$} & \multicolumn{1}{c}{$k$} & \multicolumn{1}{c}{$\bar{k}$}  \\ \cline{2-9}
  $e$ & {\cellcolor{pink!25}$e$} & \cellcolor{yellow!25}$\bar{e}$ & \cellcolor{red!25}$i$ & \cellcolor{red!25}$\bar{i}$ & \cellcolor{cyan!25}$j$ & \cellcolor{cyan!25}$\bar{j}$ & \cellcolor{green!25}$k$ & \cellcolor{green!25}$\bar{k}$ \\
  \cline{2-9}
  $\bar{e}$ & \cellcolor{yellow!25}$\bar{e}$ & \cellcolor{pink!25}$e$ & \cellcolor{red!25}$\bar{i}$ & \cellcolor{red!25}$i$ & \cellcolor{cyan!25}$\bar{j}$ & \cellcolor{cyan!25}$j$ & \cellcolor{green!25}$\bar{k}$ & \cellcolor{green!25}$k$ \\
  \cline{2-9}
  $i$ & \cellcolor{red!25}$i$ & \cellcolor{red!25}{$\bar{i}$} & \cellcolor{yellow!25}$\bar{e}$ & \cellcolor{pink!25}$e$  & \cellcolor{green!25}$k$ & \cellcolor{green!25}$\bar{k}$ & \cellcolor{cyan!25}$\bar{j}$  & \cellcolor{cyan!25}$j$ \\
  \cline{2-9}
  $\bar{i}$ & \cellcolor{red!25}$\bar{i}$ & \cellcolor{red!25}$i$ & \cellcolor{pink!25}$e$ & \cellcolor{yellow!25}$\bar{e}$ & \cellcolor{green!25}$\bar{k}$ & \cellcolor{green!25}$k$ & \cellcolor{cyan!25}$j$ & \cellcolor{cyan!25}$\bar{j}$ \\
  \cline{2-9}
  $j$ & \cellcolor{cyan!25}$j$ & \cellcolor{cyan!25}$\bar{j}$ & \cellcolor{green!25}$\bar{k}$ & \cellcolor{green!25}$k$ & \cellcolor{yellow!25}$\bar{e}$ & \cellcolor{pink!25}$e$ & \cellcolor{red!25}$i$ & \cellcolor{red!25}$\bar{i}$ \\
  \cline{2-9}
  $\bar{j}$ & \cellcolor{cyan!25}$\bar{j}$ & \cellcolor{cyan!25}$j$ & \cellcolor{green!25}$k$ & \cellcolor{green!25}$\bar{k}$ & \cellcolor{pink!25}$e$ & \cellcolor{yellow!25}$\bar{e}$ & \cellcolor{red!25}$\bar{i}$ & \cellcolor{red!25}$i$ \\
  \cline{2-9}
  $k$ & \cellcolor{green!25}$k$ & \cellcolor{green!25}$\bar{k}$ & \cellcolor{cyan!25}$j$ & \cellcolor{cyan!25}$\bar{j}$ & \cellcolor{red!25}$\bar{i}$ & \cellcolor{red!25}$i$ & \cellcolor{yellow!25}$\bar{e}$ & \cellcolor{pink!25}$e$ \\
  \cline{2-9}
  $\;\bar{k}\;$ & \cellcolor{green!25}$\;\bar{k}\;$ \hfill& \cellcolor{green!25}$\;k\;$ \hfill& \cellcolor{cyan!25}$\;\bar{j}\;$ \hfill& \cellcolor{cyan!25}$\;j\;$ \hfill& \cellcolor{red!25}$\;i\;$ \hfill& \cellcolor{red!25}$\;\bar{i}\;$ \hfill& \cellcolor{pink!25}$\;e\;$ \hfill& \cellcolor{yellow!25}$ \;\bar{e}\; $ \hfill\\
   \cline{2-9}
\end{tabular}
\label{table}
\end{table}

\subsection{Fusion rules}
\label{Frules}
The complete set of fusion rules \cite{2012NJPh...14c5024B} are presented below for the sake of completeness.

\emph{Cargeons only:}

\begin{align}
\rho_x \otimes \rho_x &= \mathds{1}, \; \   \ \rho_x \otimes \rho_y = \rho_z \\
\rho_x \otimes \Delta &= \Delta , \;  \   \ \Delta \otimes \Delta = \mathds{1} \oplus \rho_x \oplus \rho_y \oplus \rho_z \notag
\end{align}

\emph{Fluxons only:}

\begin{align}
\bar{\mathds{1}} \otimes \bar{\mathds{1}} &= \mathds{1}, \; \   \ \Phi_x \otimes \Phi_x = \mathds{1} \oplus \mathds{\bar{1}} \oplus \rho_x \oplus \bar{\rho}_x \\
\Phi_x \otimes \Phi_y &= \Phi_z \oplus \tilde{\Phi}_z , \; \   \  \bar{\mathds{1}} \otimes \Phi_x = \Phi_x \notag
\end{align}

\emph{Dyons only:}

\begin{align}
\tilde{\Phi}_x \otimes \tilde{\Phi}_x &= \mathds{1} \oplus \mathds{\bar{1}} \oplus \rho_x \oplus \bar{\rho}_x, \; \   \ \tilde{\Phi}_x \otimes \bar{\rho}_x = \tilde{\Phi}_x \\
\tilde{\Phi}_x \otimes \bar{\rho}_y &= \Phi_x , \; \   \ \bar{\rho}_x \otimes \bar{\Delta} =  \bar{\Delta} \notag\\
\bar{\Delta} \otimes \tilde{\Phi}_x &= \Sigma_x \oplus \tilde{\Sigma}_x , \; \   \ \bar{\Delta} \otimes \Sigma_x = \Phi_x \oplus \tilde{\Phi}_x \notag\\
\Sigma_x \otimes \Sigma_x &= \mathds{1} \oplus \rho_x \oplus \bar{\rho}_y \oplus \bar{\rho}_z , \;  \   \  \bar{\Sigma}_x \otimes \tilde{\Sigma}_x = \bar{\mathds{1}} \oplus \bar{\rho_x} \oplus \rho_y \oplus \rho_z \notag\\
\Sigma_x \otimes \Sigma_y &= \Phi_z \oplus \tilde{\Phi}_z \notag
\end{align}

\emph{Chargeons, fluxons and dyons:}

\begin{align}
    \rho_x \otimes \Phi_x &= \Phi_x, \; \  \ \rho_x \otimes \Phi_y = \tilde{\Phi}_y \\
    \Delta \otimes \Phi_x &= \Sigma_x \oplus \tilde{\Sigma}_x,  \; \  \ \tilde{\Phi}_x \otimes \bar{\mathds{1}} = \tilde{\Phi}_x,\notag\\
    \bar{\mathds{1}} \otimes \Sigma_x &= \tilde{\Sigma}_x,  \; \  \  \bar{\mathds{1}} \otimes \tilde{\Sigma}_x = \Sigma_x \notag\\
    \rho_x \otimes \Sigma_x &= \Sigma_x, \; \  \ \rho_y \otimes \Sigma_x = \tilde{\Sigma}_x \notag \\
    \bar{\rho}_x \otimes \Sigma_x &= \tilde{\Sigma}_x, \; \  \ \Delta \otimes \Sigma_x = \Phi_x \oplus \tilde{\Phi}_x \notag\\
    \Delta \otimes \tilde{\Sigma}_x &= \Phi_x \oplus \tilde{\Phi}_x, \; \  \ \Delta \otimes \bar{\mathds{1}} = \bar{\Delta} \notag\\
    \Phi_x \otimes \Sigma_x &= \Delta \oplus \bar{\Delta}, \; \   \ \Phi_x \otimes \Sigma_y = \Phi_z \oplus \tilde{\Phi}_z,\notag \\
    \Phi_x \otimes \Phi_x &= \mathds{1} \oplus \mathds{\bar{1}} \oplus \rho_x \oplus \tilde{\rho}_x, \; \   \ \Phi_x \otimes \Phi_y = \Phi_z \oplus \tilde{\Phi}_z,\notag\\
    \Sigma_x \otimes \Sigma_x &= \mathds{1}  \oplus \rho_x \oplus \bar{\rho}_y \oplus \bar{\rho}_z, \; \   \ \Sigma_x \otimes \Sigma_y = \Sigma_z \oplus \tilde{\Sigma}_z \notag
\end{align}

\subsection{Two qubit braid matrices}
\label{2qubit}

The two qubit braid matrices presented here can be computed with the aid of graphical calculus, given that the single qubit braid matrices are known. For a thorough discussion we refer the reader to \cite{2018QS&T....3d5004F}.

\begin{table*}\centering
\renewcommand{\arraystretch}{1.6}
\begin{center}
\caption{Values of the variables $a,b,c,d$ and $e$ when braiding $X$ and $Y$.}
\label{reptable}
\begin{tabular}{ c|c|c|c|c| } 
 \multicolumn{1}{c}{($X,Y$)} & \multicolumn{1}{c}{$\Phi_x$} & \multicolumn{1}{c}{$\Sigma_x$}\\ 
 \cline{2-3}
  $\Phi_y$ & \ \ $a=-1,\ \ \ b=e^{i \frac{\pi}{2}},\ \ \ c=e^{i \frac{3 \pi}{4}},\ \ \ d=e^{i \frac{5\pi}{4}},\ \ \ e=e^{i \frac{3\pi}{4}}$ &\ \  $a=-e^{-i \frac{\pi}{4}},\ \ \ b=e^{i \frac{\pi}{4}},\ \ \ c=1,\ \ d=-1,\ \ \ e=e^{i \frac{\pi}{2}}$ \\ 
 \cline{2-3}
 $\Sigma_y$ & $a=-e^{-i \frac{\pi}{4}},\ \ \ b=e^{i \frac{\pi}{4}}, \ \  \ c=1, \ \ d=-1,\ \ \ e=e^{i \frac{\pi}{2}}$ & $a=e^{-i \frac{\pi}{4}},\ \ \ b=e^{-i \frac{3\pi}{4}},\ \ \ c=-1,\ \ \ d=1,\ \ \ e=1$ \\ 
 \cline{2-3}
\end{tabular}
\end{center}
\end{table*}
\begin{equation}
    \sigma_1^2 (X,Y) =  \begin{pmatrix}
    a & 0 & 0 & 0 & 0 & 0 & 0 & 0\\
    0 & a & 0 & 0 & 0 & 0 & 0 & 0\\
    0 & 0 & b & 0 & 0 & 0 & 0 & 0\\
    0 & 0 & 0 & b & 0 & 0 & 0 & 0\\
    0 & 0 & 0 & 0 & a & 0 & 0 & 0\\
    0 & 0 & 0 & 0 & 0 & a & 0 & 0\\
    0 & 0 & 0 & 0 & 0 & 0 & b & 0\\
    0 & 0 & 0 & 0 & 0 & 0 & 0 & b
    \end{pmatrix}
\end{equation}

\begin{equation}
    \sigma_2^2 (X,Y) = \frac{1}{\sqrt{2}}  \begin{pmatrix}
    c & 0 & d & 0 & 0 & 0 & 0 & 0\\
    0 & c & 0 & d & 0 & 0 & 0 & 0\\
    d & 0 & c & 0 & 0 & 0 & 0 & 0\\
    0 & d & 0 & c & 0 & 0 & 0 & 0\\
    0 & 0 & 0 & 0 & c & 0 & d & 0\\
    0 & 0 & 0 & 0 & 0 & c & 0 & d\\
    0 & 0 & 0 & 0 & d & 0 & c & 0\\
    0 & 0 & 0 & 0 & 0 & d & 0 & c
    \end{pmatrix}
\end{equation}

\begin{equation}
    \sigma_3^2 (X,Y) = \frac{1}{\sqrt{2}} \begin{pmatrix}
    a & 0 & 0 & 0 & b & 0 & 0 & 0\\
    0 & b & 0 & 0 & 0 & a & 0 & 0\\
    0 & 0 & b & 0 & 0 & 0 & a & 0\\
    0 & 0 & 0 & a & 0 & 0 & 0 & b\\
    b & 0 & 0 & 0 & a & 0 & 0 & 0\\
    0 & a & 0 & 0  & b & 0 & 0 & 0\\
    0 & 0 & a & 0 & 0 & b & 0 & 0\\
    0 & 0 & 0 & b & 0 & 0 & 0 & a
    \end{pmatrix}
\end{equation}

\begin{equation}
    \sigma_4^2 (X,Y) = \frac{1}{\sqrt{2}}  \begin{pmatrix}
    c & d & 0 & 0 & 0 & 0 & 0 & 0\\
    d & c & 0 & 0 & 0 & 0 & 0 & 0\\
    0 & 0 & c & d & 0 & 0 & 0 & 0\\
    0 & 0 & d & e & 0 & 0 & 0 & 0\\
    0 & 0 & 0 & 0 & c & d & 0 & 0\\
    0 & 0 & 0 & 0  & d & e & 0 & 0\\
    0 & 0 & 0 & 0 & 0 & 0 & c & d\\
    0 & 0 & 0 & 0 & 0 & 0 & d & e
    \end{pmatrix}
\end{equation}

\begin{equation}
    \sigma_5^2 (X,Y) =  \begin{pmatrix}
    a & 0 & 0 & 0 & 0 & 0 & 0 & 0\\
    0 & b & 0 & 0 & 0 & 0 & 0 & 0\\
    0 & 0 & a & 0 & 0 & 0 & 0 & 0\\
    0 & 0 & 0 & b & 0 & 0 & 0 & 0\\
    0 & 0 & 0 & 0 & a & 0 & 0 & 0\\
    0 & 0 & 0 & 0 & 0 & b & 0 & 0\\
    0 & 0 & 0 & 0 & 0 & 0 & a & 0\\
    0 & 0 & 0 & 0 & 0 & 0 & 0 & b
    \end{pmatrix}
\end{equation}

\subsection{Exactly realizable quantum gates}
\label{exactgates}
We list below explicit forms, in terms of the elementary braid matrices, for a set of gates that can be realised exactly by braiding alone within the $\mathcal{D} (\mathds{Q}_8)$.
\begin{equation}
    S =  \begin{pmatrix}
    1 & 0\\
    0 & i
    \end{pmatrix} = \sigma_1^{-1}
\end{equation}

\begin{equation}
    {\rm H} =  \frac{1}{\sqrt{2}}\begin{pmatrix}
    1 & 1\\
    1 & -1
    \end{pmatrix} = \sigma_1 \sigma_2 \sigma_1
\end{equation}

\begin{equation}
    {\rm Pauli}-X =  \begin{pmatrix}
    0 & 1\\
    1 & 0
    \end{pmatrix} =  \sigma_2 \sigma_2
\end{equation}

\begin{equation}
    {\rm Pauli}-Y =  \begin{pmatrix}
    0 & -i\\
    i & 0
    \end{pmatrix} = \sigma_1 \sigma_1 \sigma_2^{-1} \sigma_2^{-1}
\end{equation}

\begin{align}
    {\rm Pauli}-Z =  \begin{pmatrix}
    1 & 0\\
    0 & -1
    \end{pmatrix} = \sigma_1 \sigma_1
\end{align}

\begin{equation}
    {\rm CNOT} =  \begin{pmatrix}
    1 & 0 & 0 & 0\\
    0 & 1 & 0 & 0\\
    0 & 0 & 0 & 1\\
    0 & 0 & 1 & 0
    \end{pmatrix} = \mathcal{P}\sigma_3^{-1} \sigma_4^{-1} \sigma_5^{-1} \mathcal{P}\sigma_3 \sigma_4 \mathcal{P}\sigma_3 \sigma_1 
\end{equation}

\begin{equation}
    {\rm controlled}-Z =  \begin{pmatrix}
    1 & 0 & 0 & 0\\
    0 & 1 & 0 & 0\\
    0 & 0 & 0 & 0\\
    0 & 0 & 0 & -1
    \end{pmatrix} = \sigma_1 \mathcal{P}\sigma_3^{-1} \sigma_5.
\end{equation}
The two qubit controlled gates need to be accompanied by a projective measurement $\mathcal{P}$ that projects the quantum state onto the computable subspace to avoid leakage to non-computable subspace.

\subsection{S and T matrices}
\label{SandT}
The modular S and T-matrices span the group ${\rm SL}(2,\mathds{C})$ \cite{1995hep.th...11201D}. The S-matrix can be regarded as the equivalent of a character table in the context of quantum double structure and can be computed as
\begin{equation}
    S^{AB}_{\Gamma \Lambda} = \frac{1}{H} \sum_{h_A \in C^A, h_B \in C^B} {{\rm Tr}(\Gamma(g_A^{-1} h_B g_A))}^* {\rm Tr}(\Lambda(g_B^{-1} h_A g_B))^*,
\end{equation}
where the sum is carried out over all elements belonging to the conjugacy classes $C^A$ and $C^B$ such that $[h_A,h_B] = e$ is satisfied, and $\Gamma$ and $\Lambda$ are the centralizer irreducible representations.The S-matrix is provided explicitly in  \cite{2012NJPh...14c5024B} and is provided for the sake of completeness in Table~\ref{Smat}. The T-matrix contains information about the topological spins of the particles and can be computed as
\begin{equation}
    T^{AB}_{\Gamma \Lambda} = \delta_{\Gamma,\Lambda} \delta^{A,B} e^{i 2 \pi s^A_{\Gamma}} = \frac{1}{d_{\Gamma}} {\rm Tr}(\Gamma(h^A)),
\end{equation}
where $s$ is the topological spin and $d$ is the quantum dimension.
\begin{table*}
\caption{The modular S matrix of $\mathcal{D} (\mathds{Q}_8)$. Here $\epsilon_{ij}=2 \delta_{ij} -1$ and $\delta_{ij}$ is the Kronocker delta.}
\setlength{\tabcolsep}{10pt}
\renewcommand{\arraystretch}{1.5}
\label{Smat}
\begin{tabular}{c|c|c|c|c|c|c|c|c|c|c|}

\multicolumn{1}{c}{$S$}  & \multicolumn{1}{c}{$\mathds{1}$}  & \multicolumn{1}{c}{$\bar{\mathds{1}}$}  & \multicolumn{1}{c}{$\rho_j$}  & \multicolumn{1}{c}{$\bar{\rho_j}$} & \multicolumn{1}{c}{$\Delta$} & \multicolumn{1}{c}{$\bar{\Delta}$} &\multicolumn{1}{c}{ $\Phi_j$} & \multicolumn{1}{c}{$\tilde{\Phi}_j$}  & \multicolumn{1}{c}{$\Sigma_j$}  &\multicolumn{1}{c}{ $\tilde{\Sigma}_j$}  \\ \cline{2-11}

$\mathds{1}$  & $\frac{1}{8}$ & $\frac{1}{8}$ & $\frac{1}{8}$ & $\frac{1}{8}$ & $\frac{1}{8}$ & $\frac{1}{4}$ & $\frac{1}{4}$ & $\frac{1}{4}$ & $\frac{1}{4}$ & $\frac{1}{4}$ \\ \cline{2-11}

$\bar{\mathds{1}}$  & $\frac{1}{8}$ & $\frac{1}{8}$ & $\frac{1}{8}$ & $\frac{1}{8}$ & $-\frac{1}{4}$ & $-\frac{1}{4}$ & $\frac{1}{4}$ & $\frac{1}{4}$ & $-\frac{1}{4}$ & $-\frac{1}{4}$ \\ \cline{2-11}

$\rho_i$  & $\frac{1}{8}$ & $\frac{1}{8}$ & $\frac{1}{8}$ & $\frac{1}{8}$ & $\frac{1}{4}$ & $\frac{1}{4}$ & $\frac{1}{4}\epsilon_{ij}$ & $\frac{1}{4}\epsilon_{ij}$ & $\frac{1}{4}\epsilon_{ij}$ & $\frac{1}{4}\epsilon_{ij}$  \\ \cline{2-11}

$\bar{\rho_i}$  & $\frac{1}{8}$ & $\frac{1}{8}$ & $\frac{1}{8}$ & $\frac{1}{8}$ & $-\frac{1}{4}$ & $-\frac{1}{4}$ & $-\frac{1}{4}\epsilon_{ij}$ & $-\frac{1}{4}\epsilon_{ij}$  & $\frac{1}{4}\epsilon_{ij}$ & $\frac{1}{4}\epsilon_{ij}$  \\ \cline{2-11}

$\Delta$  & $\frac{1}{4}$ & $-\frac{1}{4}$ & $\frac{1}{4}$ & $-\frac{1}{4}$ & $\frac{1}{2}$ & $-\frac{1}{2}$ & $0$ & $0$ & $0$ & $0$ \\ \cline{2-11}

$\bar{\Delta}$  & $\frac{1}{4}$  & $-\frac{1}{4}$ & $\frac{1}{4}$ & $-\frac{1}{4}$ & $-\frac{1}{2}$ & $\frac{1}{2}$ & $0$ & $0$ & $0$ & $0$ \\ \cline{2-11}

$\Phi_i$  & $\frac{1}{4}$ & $\frac{1}{4}$ & $\frac{1}{4}\epsilon_{ij}$ & $-\frac{1}{4}\epsilon_{ij}$ & $0$ & $0$ & $\frac{1}{2}\delta_{ij}$ & $-\frac{1}{2}\delta_{ij}$ & $0$ & $0$ \\ \cline{2-11}

$\tilde{\Phi}_i$  & $\frac{1}{4}$ & $\frac{1}{4}$ & $\frac{1}{4}\epsilon_{ij}$ & $-\frac{1}{4}\epsilon_{ij}$ & $0$ & $0$ & $-\frac{1}{4}\delta_{ij}$ & $\frac{1}{4}\delta_{ij}$ & $0$ & $0$ \\ \cline{2-11}

$\Sigma_i$  & $\frac{1}{4}$  & $-\frac{1}{4}$ & $\frac{1}{4}\epsilon_{ij}$ & $\frac{1}{4}\epsilon_{ij}$ & $0$ & $0$ & $0$ & $0$ & $\frac{1}{4}\delta_{ij}$ & $-\frac{1}{4}\delta_{ij}$  \\ \cline{2-11}

$\tilde{\Sigma}_i$ & $\frac{1}{4}$ & $-\frac{1}{4}$ & $\frac{1}{4}\epsilon_{ij}$ & $\frac{1}{4}\epsilon_{ij}$ & $0$ & $0$ & $0$ & $0$ & $-\frac{1}{4}\delta_{ij}$  & $\frac{1}{4}\delta_{ij}$ \\ \cline{2-11}
\end{tabular}
\end{table*}

\subsection{$F$ and $R$ symbols}
\label{RandF}

In this section we provide the background material required to work out the braid matrices. The single qubit braid matrices are given by $\sigma_1 = R$ and $\sigma_2 = F^{-1} R F$, where $R$ and $F$ correspond to the anyon interchange and change of fusion basis, respectively, and are given by \cite{1995hep.th...11201D}

\begin{equation}
   R_{j_i j_j}^{j_k} = \sum_{m_i, m_j} \sum_{m_q, m_p} \sigma_{m_i,m_q}^{m_j,m_p} \circ \mathcal{R}_{j_i, j_j}^{(m_i,m_q,),(m_j,m_p)}
\end{equation}
and
\begin{align}
\noindent
\label{Fsymb}
 &[F^{j_q}_{j_i,j_j,j_k}]_{j_l}^{j_p} = \sum_{m_i,m_j,m_k,m_q,m_p} \left[ \begin{array}{cc|c}
    j_{i} & j_{j} & j_{l} \\
    m_{i} & m_{j} & m_{l} \\
    \end{array}
    \right]\left[ \begin{array}{cc|c}
    j_{l} & j_{k} & j_{q} \\
    m_{l} & m_{k} & m_{q} \\ 
    \end{array}  
    \right] 
    \times
    \nonumber \\ & \left[ \begin{array}{c|cc}
    j_{q} & j_{p} & j_{i} \\
    m_{q} & m_{p} & m_{i} \\
    \end{array}
    \right]\left[ \begin{array}{c|cc}
    j_{p} & j_{j} & j_{k} \\
    m_{p} & m_{j} & m_{k} \\
    \end{array}
    \right],
\end{align}
where $j_i$ and $m_i$ are the topological spins and magnetic moments, respectively, and the brackets denote the quantum double Clebsch--Gordan coefficients which are given by Eqs. \eqref{CG}-\eqref{offdiagelem}. The $\sigma_{m_i,m_q}^{m_j,m_p}$ are elements of the permutation operator $\sigma$ which is equivalent to a decoupling followed by a recoupling where the anyons are swapped, i.e.
\begin{equation}
 \sigma_{m_i,m_q}^{m_j,m_p} = \left[ \begin{array}{cc|c}
    j_{i} & j_{j} & j_{k} \\
    m_{i} & m_{j} & m_{k} \\
    \end{array}
    \right] \left[ \begin{array}{c|cc}
    j_{k} & j_{j} & j_{i} \\
    m_{k} & m_{p} & m_{q} \\
    \end{array}
    \right]
\end{equation}
and the $\mathcal{R}_{j_i, j_j}^{(m_i,m_q,),(m_j,m_p)}$ elements are given by
\begin{equation}
   \mathcal{R}_{j_i, j_j}^{(m_i,m_q,),(m_j,m_p)} = \sum_h \sum_g \Lambda_{m_i, m_q}^{j_i}(P_g e) \otimes \Lambda_{m_j, m_p}^{j_j}(P_h g),
\end{equation}
where $\Lambda^{j_i}_{m_i,m_q}$ are the representations corresponding to the topological charge $j_i$ mapping the quantum double element $P_h g$ (a gauge transformation $g$ followed by a flux measurement $P_h$) to a matrix implementing the quantum double action. The Clebsch--Gordan coefficients can be derived analytically by unpacking the representations via the projection operators in the representation theory of the quantum double. In doing so, one finds that the coefficients must satisfy 
\begin{align}
\label{CG}
   &\sum_n \left[ \begin{array}{cc|c}
    j_{i} & j_{j} & j_{l} \\
    m_{i} & m_{j} & m_{l} \\
    \end{array}
    \right]_ n^* \left[ \begin{array}{cc|c}
    j_{q} & j_{p} & j_{k} \\
    m_{q} & m_{p} & m_{k} \\
    \end{array}
    \right]_n = \notag \\
   & \frac{d_{j_k}}{|H|} \sum\limits_{h, g} \Lambda^{j_k}_{m_k m_l} (P_h g)^* \sum\limits_{h' h'' = h} \Lambda^{j_i}_{m_i m_q}(P_{h'}g) \Lambda^{j_j}_{m_j m_q}(P_{h''} g),
\end{align}
where $n$ is the multiplicity of the corresponding irreducible representation. In the $\mathcal{D}(\mathds{Q}_8)$ anyon model all fusion outcomes have unit multiplicity meaning that we can solve Eq. \eqref{CG} analytically since there is only one term on the left hand side of the equation. Setting $i=q$, $j=p$ and $k=l$ we find the solution corresponding to the diagonal elements of the representations
\begin{align}
\label{diagelem}
    &\left[ \begin{array}{cc|c}
    j_{i} & j_{j} & j_{k} \\
    m_{i} & m_{j} & m_{k} \\
    \end{array}
    \right] = \notag \\
    &\sqrt{\frac{d_{j_k}}{|H|} \sum\limits_{h, g} \Lambda^{j_k}_{m_k m_k} (P_h g)^* \sum\limits_{h' h'' = h} \Lambda^{j_i}_{m_i m_i}(P_{h'}g) \Lambda^{j_j}_{m_j m_j}(P_{h''} g)}.
\end{align}

Finally, we can divide Eq.~\eqref{CG} by the solution given by Eq.~\eqref{diagelem} to obtain the full solution
\begin{align}
\label{offdiagelem}
    &\left[ \begin{array}{cc|c}
    j_{q} & j_{p} & j_{k} \\
    m_{q} & m_{p} & m_{k} \\
    \end{array} \notag
    \right]_{(m_i,m_j,m_k)} = \\
    &\sqrt{\frac{d_{j_k}}{|H|}}\frac{ \sum\limits_{h, g} \Lambda^{j_k}_{m_k m_l} (P_h g)^* \sum\limits_{h' h'' = h} \Lambda^{j_i}_{m_i m_q}(P_{h'}g) \Lambda^{j_j}_{m_j m_q}(P_{h''} g)}{\sqrt{ \sum\limits_{h, g} \Lambda^{j_k}_{m_k m_k} (P_h g)^* \sum\limits_{h' h'' = h} \Lambda^{j_i}_{m_i m_i}(P_{h'}g) \Lambda^{j_j}_{m_j m_j}(P_{h''} g)}}.
\end{align}

This result is similar to that obtained with a different method in \cite{2006CoPhC.174..903R} for regular finite groups. One can recover Eq.~\eqref{offdiagelem} from their derivation by considering the quantum double of the discrete group.

\end{appendix}

\end{document}

%% file: output.bbl
%